\documentclass[letterpaper,conference]{IEEEtran}
\IEEEoverridecommandlockouts
\usepackage{cite}
\usepackage{tabularx}
\usepackage{textcase}
\usepackage{amsmath,amssymb,amsfonts}
\usepackage{algorithmic}
\usepackage{graphicx}
\usepackage{textcomp}
\usepackage{xcolor}
\usepackage{booktabs, tabularx}
\usepackage{rotating}
\usepackage{comment}
\usepackage{placeins}
\usepackage{hyperref}
\usepackage{textcomp}
\usepackage{amsfonts} 
\usepackage{float}
\usepackage{array}
\usepackage{mathtools}
\usepackage{multirow}
\usepackage{pifont}

\usepackage[top=54pt, left=46pt, right=46pt, bottom=75pt]{geometry}

\newcommand{\xmark}{\ding{55}}

\usepackage[caption=false]{subfig}
\usepackage{stfloats}
\def\BibTeX{{\rm B\kern-.05em{\sc i\kern-.025em b}\kern-.08em
		T\kern-.1667em\lower.7ex\hbox{E}\kern-.125emX}}

\usepackage{array}
\newcolumntype{M}[1]{>{\centering\arraybackslash}m{#1}}
\newcolumntype{N}{@{}m{0pt}@{}}
\usepackage{makecell}

\begin{document}

	\title{From C-Band to mmWave-Band: Ray-Tracing-Assisted 5G-Based Indoor Positioning in Industrial Scenario \\
	}
	
	\makeatletter
    \newcommand{\linebreakand}{%
    \end{@IEEEauthorhalign}
    \hfill\mbox{}\par
    \mbox{}\hfill\begin{@IEEEauthorhalign}
}
\makeatother

	\author{
    \IEEEauthorblockN{
        Karthik Muthineni\IEEEauthorrefmark{1}\IEEEauthorrefmark{2}, Alexander Artemenko\IEEEauthorrefmark{1}
    }
    \IEEEauthorblockA{\IEEEauthorrefmark{1} Corporate Sector Research and Advance Engineering, Robert Bosch GmbH, Renningen, Germany}
    \IEEEauthorblockA{\IEEEauthorrefmark{2} Department of Signal Theory and Communications, Universitat Politècnica de Catalunya (UPC), Barcelona, Spain}
    \IEEEauthorblockA{Email: \IEEEauthorrefmark{1}$\{${karthik.muthineni, alexander.artemenko}$\}$@de.bosch.com}
}

	
		
 
	\maketitle
	
	\begin{abstract}
            Private fifth-generation (5G) networks are increasingly becoming the industry's choice of wireless communication networks for accelerating production processes. In this context, the role of 5G in providing precise positioning services in indoor industrial scenarios has also been actively discussed. However, the achievable indoor positioning accuracy depends on the radio propagation conditions persisting in the scenario. In this paper, using a Ray-Tracing (RT) engine, we investigate the radio environment in C-band (3.775 GHz) as well as the mmWave-band (26.85 GHz) for a detailed 3D geometric model of the dense-clutter industrial production hall under different emulation setups and categorize the dominant Non-Line-of-Sight (NLoS) MultiPath Components (MPCs). We then evaluate the achievable Observed Time Difference of Arrival (OTDoA) based positioning accuracy in the C-band and the mmWave-band by computing the position of User Equipment (UE) using only first-arriving MPCs.

		
	\end{abstract}
	
	
	\section{Introduction}
	Industry 4.0 envisions a digitally connected industry with accurate and precise positioning technology that allows mobile robots like Automated Guided Vehicles (AGV) to navigate autonomously and provide Positioning-as-a-Service (PaaS) to other manufacturing systems. For instance, the manufacturing robot can use the PaaS to request the current position of the AGV and place the items ready for it to pick up, minimizing the wait time and improving efficiency~\cite{BOSCH}. Such applications demand real-time data with strict requirements on availability, reliability, latency, and positioning accuracy.
 
    The roll-out of fifth-generation (5G) cellular communication technology has attracted the attention of industries due to its guaranteed Quality of Service (QoS) and integrated positioning capabilities. Till today, the adoption of cellular technology in industries is not seen widely because it acts in a licensed spectrum and mandates a license from the national telecom regulator. Although the same trend follows in 5G, industries can now have diverse options for the deployment of the network on their premises. Such a type of network is known to be a private network or Non-Public Network (NPN) in terms of the Third Generation Partnership Project (3GPP) terminology~\cite{3GPP}.

    A private network operation depends on the deployment model and the chosen radio spectrum. The network can be deployed as a standalone or in conjunction with the public network, where the integration model can follow one out of the three possible configurations identified by the 5G Alliance for Connected Industries and Automation (5G-ACIA) \cite{5GACIA}. On the other hand, the radio spectrum for private 5G networks is available in different types. First, a licensed spectrum, which is attractive to Mobile Network Operators (MNO). Second, the unlicensed spectrum, which is free of charge and shared among different entities~\cite{STRINATI2020}. In addition, several countries started allocating dedicated spectra for enterprises. For example, the German government had reserved a 100~MHz spectrum in the C-band (3.7-3.8~GHz) and another 1~GHz spectrum in the mmWave-band (26.5-27.5~GHz) for industrial applications~\cite{BND}.

    To extend the 5G footprint beyond public networks and to support the requirements of industry verticals, 3GPP has targeted the positioning accuracy of $<$3~m for indoor deployments in release~16. More demanding requirements with an accuracy of a few decimeters inside the industry are suggested in release~17. Achieving the 3GPP's needs in the industrial setting becomes challenging, where the Non-Line-of-Sight (NLoS) condition occurs frequently, giving rise to Multi-Path Components (MPCs). These MPCs impact the accuracy of time and/or angle estimation of the first arriving MPC, leading to outliers. Position estimation with outliers in the measurements leads to inaccurate results and in such cases new approaches for eliminating the outliers become essential. However, in this work, we only focus on the initial positioning, which refers to the implementation of state-of-the-art techniques for position computation. This leads to the research question, that motivates this work. What accuracy level can be obtained in the specific industrial scenario by using 5G for the initial positioning?

    The main contributions of this paper are:
    \begin{itemize}
        \item Using a Ray-Tracing (RT) engine, we analyze the radio channel in the C-band (3.775~GHz) as well as in the mmWave-band (26.85~GHz) and categorize the distribution of the first-arriving MPC, for a specific industrial scenario.
        \item We analyze the achievable Observed Time Difference of Arrival (OTDoA) positioning accuracy in C-band and mmWave-band by using first-arriving MPC as the preferred link of choice to compute the User Equipment (UE) position~\cite{JATIVA2002}, located in a specific industrial scenario.
    \end{itemize}

	\section{Analytical Model for OTDoA-Based Indoor Positioning}
        \label{Analytical Model for 5G-Based Indoor Positioning}
        In this section, we describe the analytical model for 5G-based positioning using the OTDoA, which is a UE-based network-assisted positioning technique. In OTDoA, multiple Base Stations (BSs) transmit reference signals and the UE measures the difference in Time of Arrival (ToA) of the reference signals. 
        

        Suppose a UE with an omnidirectional antenna is located at the point $\{p_{ue} = (x_{ue} \,, y_{ue})\,,\,p_{ue} \in \mathbb{R}^2\}$ in a 2-dimensional (2D) space. The 5G positioning infrastructure includes $\mathrm{K}$ BSs each with an omnidirectional antenna whose positions are known to the UE. Let $B = \{1\,,\,2\,,\,.\,.\,.\,,\,\mathrm{K}\}$ be the indices of the BSs located at the points $\{p_{bs} = (x_{bs} \,, y_{bs})\,, \,p_{bs} \in \mathbb{R}^2\,, \,bs \in B\}$. We assume that the BS clocks are synchronized, but are not synchronized with the UE clock. Moreover, the assumed attributes about the 5G positioning infrastructure are listed in Table~\ref{tab1}.
        
        Let $LS = \{1\,,\,2\,,\,.\,.\,.\,,\,\mathrm{G}\}$ be the set of BSs that are in Line-of-Sight (LoS) with the UE. On the other hand, let $NLS (= \mathrm{H})$ be the set of BSs that are in NLoS with $\mathrm{H} = \mathrm{K} - \mathrm{G}$. The radio signal transmitted by the BS might reach the UE in more than one possible path due to the NLoS condition. The received signal in a multipath NLoS environment from the $\textrm{BS}_{bs}$ can be defined as

        \begin{equation}
            s_{bs}(t) = \sum_{i=1}^{N_{bs}} A_{bs}^i \cdot r(t - \tau_{bs}^i), \,\,\,\,\,bs \in B
            \label{eq1}
        \end{equation}
        where $r(t)$ represents the transmitted reference signal, $N_{bs}$ is the number of MPCs received from the $\textrm{BS}_{bs}$, $A_{bs}^i$ and $\tau_{bsi}^i$ are the amplitude and ToA of the $i$-th MPC of the $s_{bs}(t)$. In this work, we use the first-arriving signal path as the preferred link of choice to compute the position of the UE. Therefore, the ToA of the first-arriving signal path is defined as

        \begin{equation}
            \tau_{bs}^1 = \frac{1}{c}\left\{\sqrt{(x_{ue} - x_{bs})^2 + (y_{ue} - y_{bs})^2} + pl_{bs}^1 \right\}, \,\,\,\,\,bs \in B
            \label{eq2}
        \end{equation}
        where $c$ is the speed of light and $pl_{bs}^1$ is the path length error, induced due to the extra distance traveled by the NLoS MPC. However, if the BS is in LoS with the UE, the path length error will be zero, i.e., $\{pl_{bs}^1 = 0\,, \,bs \in LS\}$. By taking arbitrary BS as a reference $e$, the $\mathrm{K} - 1$ TDoAs  of the first-arriving signal path are computed as

        \begin{equation}
            TDoA_{ke}^1 = |(\tau_{k}^1 - \tau_{e}^1) - \delta_{ke}|, \,\,\,\,\,k \in B \setminus \{r\}
            \label{eq3}
        \end{equation}
        where $\delta_{ke}$ represents the time period between the two BS transmissions. To compute the position of the UE, (\ref{eq3}) can be converted to a difference in the distance $\Delta{d}$ by multiplying with speed of light \textit{c}, which yields a non-linear equation of the form

         \begin{equation}
         \begin{split}
                (\Delta{d})_{ke}^1 = \Big[\sqrt{(x_{ue}-x_k)^2 + (y_{ue}-y_k)^2}\\-\sqrt{(x_{ue}-x_e)^2 + (y_{ue}-y_e)^2}\Big].
                \label{eq4}
        \end{split}
        \end{equation}

        The UE position $(x_{ue}\,,\,y_{ue})$ is computed by solving the set of $\mathrm{K} - 1$ equations for BSs using the Non-Linear Least Squares (NLS) method. At least, three BSs are required to obtain the position of UE in 2D.








  \begin{table}[b]
    \caption{Assumed attributes of the 5G positioning infrastructure.}
    \begin{center}
    {\renewcommand{\arraystretch}{1.4}
    \begin{tabular}[t]{ | p{5.2cm} | p{2.35cm}| } 
    \hline
    \hfil Order of Base Station (BS) transmissions & \hfil $\textrm{BS}_1$\,,\,$\textrm{BS}_2$\,,\,.\,.\,.\,,\,$\textrm{BS}_\mathrm{K}$ \\ 
    \hline
    \hfil Precision of synchronization among BSs & \hfil 10 ns \\ 
    \hline
    The time period between BS transmissions ($\delta$) & \hfil 10 ms\\ 
    \hline
    \end{tabular}
    }
    \end{center}
    \label{tab1}
    \end{table}

        \section{Industrial Scenario for 5G-Based Indoor Positioning}
        In this section, we describe the geometric model of the industrial scenario, and the channel emulation settings used for RT are presented.
        
        \begin{table}
         \caption{Channel Emulation settings defined in the Winprop.}
         \begin{center}
         {\renewcommand{\arraystretch}{1.4}
            \begin{tabular}[b]{ | p{3.4cm} | p{1.5cm}| p{1.5cm} |} \hline
                
                \hfil Multiple access & \multicolumn{2}{|c|}{OFDM}\\ \hline
                \hfil Duplex separation & \multicolumn{2}{|c|}{TDD}\\ \hline
                \hfil Center frequency & \multicolumn{2}{|c|}{3.775 GHz / 26.85 GHz}\\
                 \hline
                \hfil Bandwidth & \multicolumn{2}{|c|}{100 MHz / 400 MHz}\\ \hline
                \hfil Sub-carrier spacing & \multicolumn{2}{|c|}{30 KHz / 120 KHz}\\ \hline
                \centering Base Station (BS) and User Equipment (UE) antennas & \multicolumn{2}{|c|}{Omnidirectional}\\ \hline
                \hfil BS transmit power & \multicolumn{2}{|c|}{20 dBm}\\ \hline
                \hfil Emulation time & \multicolumn{2}{|c|}{60 s}\\ \hline
                \hfil Propagation & \hfil Direct ray & \hfil \ding{51}\\ \cline{2-3}
                     & \centering Penetration & till 2nd order\\ \cline{2-3}
                     & \centering Reflection & till 2nd order\\ \cline{2-3}
                     & \centering Diffraction & till 1st order\\ \cline{2-3}
                     & \centering Scattering & \hfil \xmark\\ \hline \cline{2-3} 
            \end{tabular}
        }
        \end{center}
        \label{tab2}
        \end{table}
        \subsection{Geometric Model}
        The industrial scenario considered in this paper corresponds to a production hall of Bosch located in Blaichach, Germany, that has dimensions of 42~m $\times$ 46~m $\times$ 8.8~m, enclosed with thick walls, concrete ground, and a high ceiling. The production hall includes three machine lines with manufacturing robot arms, an open area for storing plastic containers, and steel pipes. The physical scenario is scanned with the 3D laser scanners to generate a point cloud, followed by conversion to the CAD model, from which the RT model has been built. A large number of objects in the production hall are considered to be metals. Besides, the metallic objects, the objects made up of concrete, plastic, wood, and glass materials are also included in the RT model, and the properties of these materials are defined according to the International Telecommunication Union (ITU) recommendation~\cite{ITU2015}. We point the interested reader to~\cite{HAN2022}, for the entire process chain of building an RT model from the 3D point cloud data.
        
        The top-down view of an indoor environment of the RT-enabled production hall is depicted in Fig.~\ref{fig: 1a}. Furthermore, Fig.~\ref{fig: 1b} depicts the 3D model with a wide view from one of the BSs. The Area-of-Interest (AoI) for indoor positioning spans roughly 29~m $\times$ 25~m and the 5G infrastructure arrangement comprises four BSs with a height of 4~m positioned at the four corners of AoI. In this work, we define 23 Points-of-Interest (PoI) with a height of 1~m, represented by red squares as shown in Fig.~\ref{fig: 1a}. These PoIs constitute the positions of the UEs (AGVs in this paper), whose positions need to be determined. At these PoIs, the positioning accuracy is evaluated by computing the 2D positioning error with respect to their ground truths. The performance of our algorithm is investigated under two different emulation setups. The first setup corresponds to the static scenario, where, no dynamic movement is considered in the production hall. The second setup corresponds to be dynamic, where, for the same BS deployment the PoIs remain static but in the background environment the dynamic movement of forklifts with the dimensions of 3~m $\times$ 1.2~m $\times$ 2.5~m, moving at a speed of 1~m/s (3.6~km/h) are considered. This introduces randomness in the ToA measurements due to the NLoS condition that occurs when a forklift enters and leaves the site in between the BS and PoI.
         \subsection{Channel Emulation} 
         
        The radio coverage planning software Altair Winprop\footnote{https://web.altair.com/winprop-telecom} is used for performing RT on top of emulation setups with reasonable parameters as listed in Table \ref{tab2}. The propagation mechanisms considered in this paper include the LoS path, penetrations up to the second order that can occur when the radio wave interacts with the objects made with glass material, reflections up to the second order, and diffraction up to the first order. The diffuse scattering due to the object roughness is ignored, because the objects in the production hall are smoother. Based on the geometry model of the scenario and the channel emulation settings defined in the Winprop, the ToA measurements are generated. Furthermore, these measurements are utilized by the positioning algorithm for locating the UE as described in Section \ref{Analytical Model for 5G-Based Indoor Positioning}.  
        \begin{figure}[t]
            \centering
            \captionsetup{width=0.85\linewidth}
            \subfloat[The two-dimensional (2D) model of the production hall with four Base Stations (BS) and 23 ground truth Points-of-Interest (PoI) (red squares). The black rectangles with the arrow mark represent the forklifts and their moving directions.]{
            \includegraphics[clip, angle=-90, trim=3cm 10.88cm 4cm 4cm, width=0.315\textwidth]{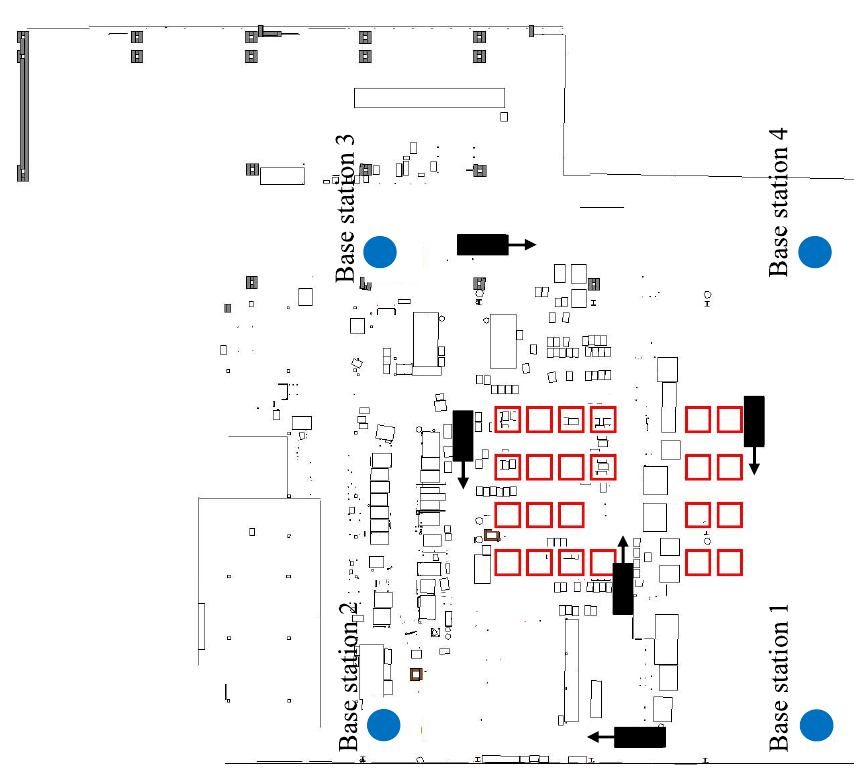}
            \label{fig: 1a}
            }
            
            \subfloat[The three-dimensional (3D) model of the production hall with a wide view from one of the BS.]{
            \includegraphics[clip, trim=0cm 0cm 0cm 0cm, width=0.4\textwidth]{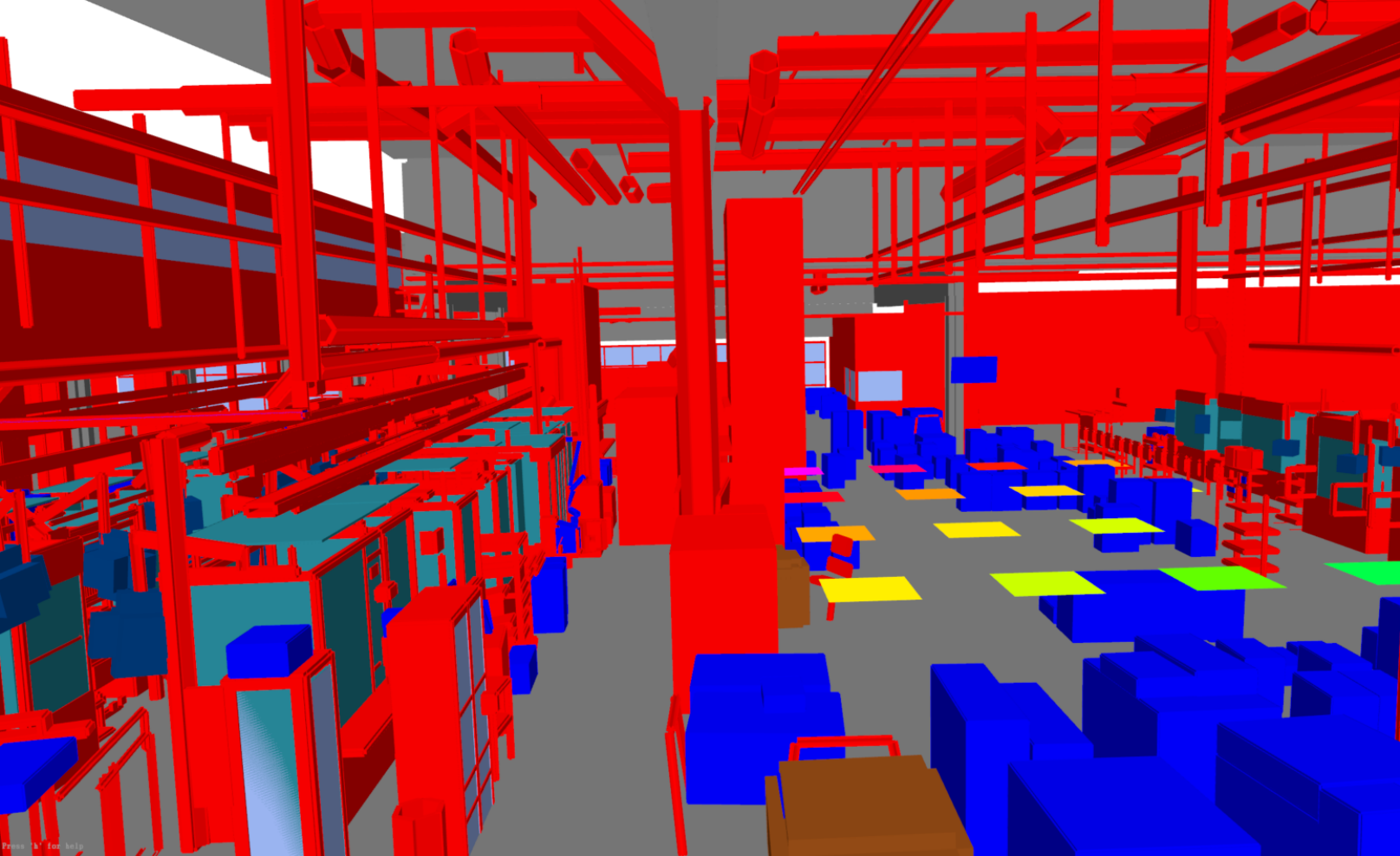}%
            \label{fig: 1b}
            }
            \caption{The geometric model of the industrial scenario.}
        \end{figure}

        \section{Results and Discussion}
        In this section, we describe and present the distribution of first-arriving MPC being LoS, penetrated, diffracted, or reflected. Moreover, the 2D positioning results obtained in our emulation setups are provided.

        \subsection{Radio channel characterization}
        Analyzing the radio channel is essential for understanding the positioning results obtained in the NLoS multipath environment because the extra path lengths induced by the MPCs are the main source of the positioning error. We carry out the RT emulation in our static and dynamic setups. Thereafter, we categorize the first-arriving MPC from each BS to the UEs into four categories, namely, LoS, penetration (through glass obstacles), diffraction (at the edges of the obstacles), reflection, and second-order interactions which involves a combination of these effects. Analyzing the distribution plots in Fig. \ref{fig:2}, one can note that under both static and dynamic setups the majority of first-arriving MPCs in C-band as well as in mmWave-band belong to the category of either LoS or diffraction. Therefore, the diffracted paths are the leading NLoS components in our emulation setups with few penetrations, reflections, and other second-order interactions. It is to be noted that the total number of first-arriving MPCs between 4 BSs and 23 PoIs should be 92. However, few PoIs receive more than 1 MPC, which corresponds to the first arrival. Therefore, the distribution of first-arriving MPCs depicted in Fig. \ref{fig:2} involves a higher number.


         \begin{figure}[t]
            \centering
            \captionsetup{width=0.94\linewidth}
            \subfloat[Static emulation setup.]{
            \includegraphics[trim={8.5cm 1cm 4cm 5cm}, clip, width=0.45\textwidth]{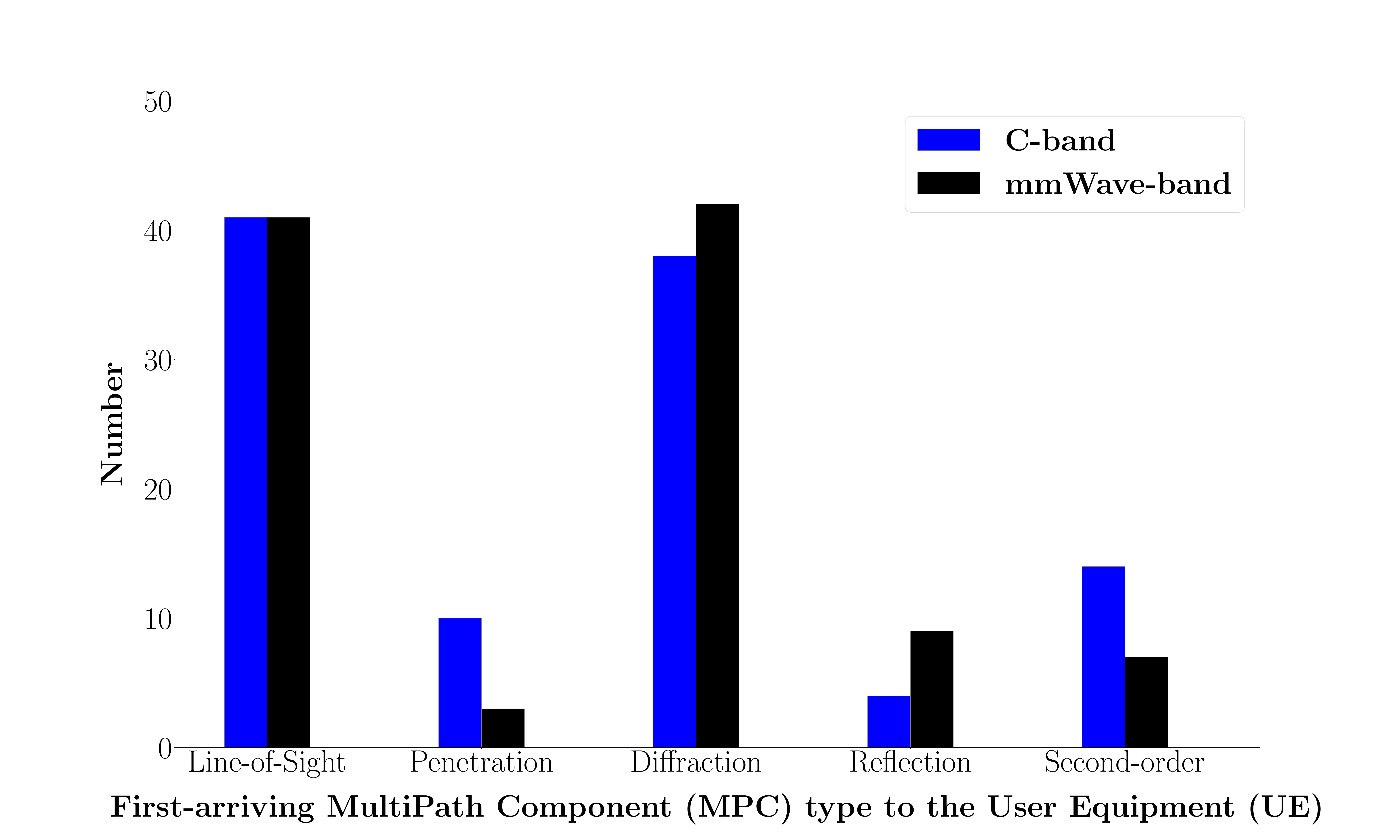}
            \label{fig:2a}
            }
            
            \subfloat[Dynamic emulation setup.]{
            \includegraphics[trim={8.5cm 1cm 4cm 5cm}, clip, width=0.45\textwidth]{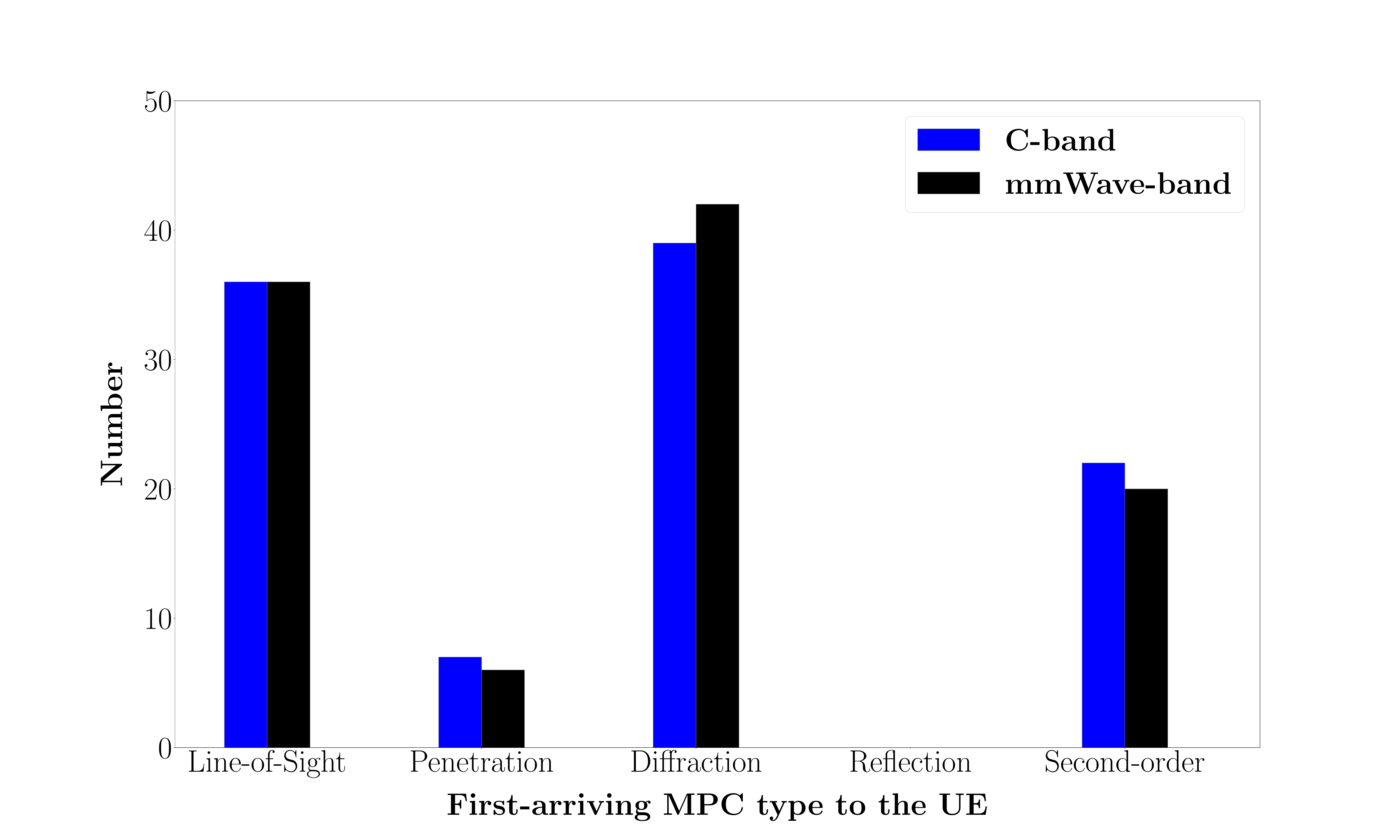}%
            \label{fig:2b}
            }
            \caption{Distribution of first-arriving MultiPath Components (MPCs).}
            \label{fig:2}
        \end{figure}

        \subsection{Indoor positioning performance}
        The 2D positioning results obtained in the static and dynamic emulation setups are depicted in Fig. \ref{fig:3}. The red dots represent the positions of the indoor 5G BSs. The squares designate the ground-truth positions of PoIs and the color associated with each PoI represents the 2D positioning error. Table \ref{tab3} summarizes the positioning error obtained in the investigated emulation setups for different frequency bands w.r.t. average error and percentile.

        \subsubsection{Static emulation setup}
        Analyzing the positioning error per PoI for C-band in Fig. \ref{fig:3a}, one can note that few PoIs are colored in black indicating very low positioning error achieved for UEs at these locations. The first-arriving MPCs to these PoIs are either LoS or diffracted rays that have similar path lengths to that of the LoS path. There are other PoIs with high positioning errors, as these are located behind machines around the pillars of the production hall. Due to the absence of LoS links from the BSs, these PoIs use first-arriving NLoS MPC to compute the position. The extra path length error induced by the NLoS MPC, makes the UE misjudge the actual distance to the BS, causing distance estimation error and leading to poor positioning performance. Similar behavior can be observed for mmWave-band as shown in Fig. \ref{fig:3b}. However, for some PoIs, a slight increase in positioning error can be observed. For instance, let's compare the obtained positioning error for PoI~16 in C-band and mmWave-band. In C-band, the first-arriving MPC from the BS~3 corresponds to the scattered ray with two interactions with the first interaction being a reflection from the ceiling indicated by S1 and the second interaction being penetration through a glass obstacle indicated by S2 as shown in Fig. \ref{fig:4a}. On the other hand, for the same BS-PoI pair, in mmWave-band, the first-arriving MPC corresponds to the scattered ray also with two interactions with the first interaction being a diffraction from the column of a building indicated by S1 and the second interaction being penetration through a glass obstacle indicated by S2 as shown in Fig. \ref{fig:4b}. Due to the larger distance covered by the scattered ray involving diffraction as well as penetration before reaching the PoI, the induced path length error is large compared to the path length error induced by the scattered ray involving reflection and penetration in C-band. This leads to inaccurate position estimation and that explains the increase in positioning error for PoI 16 in mmWave-band. Even though the BS and the PoI are not moving relative to each other, the obtained position estimation result differs when the operating frequency is changed from C-band to the mmWave-band. This is because of the presence of obstacles with different material properties in the scenario, which has a severe impact on signal propagation and limits the number of MPCs reaching PoI in the mmWave-band. As a result, the ray that corresponds to the first-arriving MPC in the mmWave-band might not be the same first-arriving MPC in the C-band as well. 
        
        The Cumulative Distribution Function (CDF) of 2D positioning error is shown in Fig. \ref{fig:5}. Looking at the curves of both the C-band and mmWave-band for a static setup, one can note that, the PoIs achieve similar performance having an error of around 1.2~m at the 90th percentile. For the current setup, as there are no dynamics involved, the difference in the positioning performance between the C-band and the mmWave-band can be seen only for a few PoIs and at the higher percentiles of the CDF function. The position estimation accuracy of PoIs located behind the machines suffers the most. For such PoIs, fewer LoS paths are present, and the available NLoS paths are longer in length. It is noticed that around the 95th percentile, C-band has around 9$\%$ smaller errors than the mmWave-band. Limiting our deployment and analysis to only four BSs, we were able to achieve average positioning errors of 0.34~m and 0.35~m in the C-band and mmWave-band respectively. 

        \begin{table}[hbt!]
            \caption{Summary of 2D positioning error.}
            \begin{center}
            \newcommand\T{\rule{0pt}{2.6ex}}       
            \newcommand\B{\rule[-3.8ex]{0pt}{0pt}} 
            {\renewcommand{\arraystretch}{1.4}
                \begin{tabular}[b]{|M {2.3cm} |M {1.2cm} |M {1.3cm} |M {1.4cm} |}
                    \hline
                     & {\textbf{Setup}} & {\hfil{\textbf{\vtop{\hbox{\strut Frequency}\hbox{\strut band}}}}\B} & {\hfil {\textbf{TDoA NLS}}} \\ 
                    \hline
                     
                     \multirow{2}{*}{\textbf{\vtop{\hbox{\strut Average positioning}\hbox{\strut error [m]}}}} &  \multirow{2}{*}{Static} & \hfil C-band & \hfil 0.34 \\ \cline{3-4}
                     & & \hfil mmWave & \hfil 0.35\\ \cline{2-4}
                     & \multirow{2}{*}{Dynamic} & \hfil C-band & \hfil 0.43 \\ \cline{3-4}
                     & & \hfil mmWave & \hfil 0.46 \\ \cline{1-4}
                     \hline

                     \multirow{2}{*}{\textbf{90th percentile [m]}} &  \multirow{2}{*}{Static} & \hfil C-band & \hfil 1.2 \\ \cline{3-4}
                     & & \hfil mmWave & \hfil 1.2\\ \cline{2-4}
                     & \multirow{2}{*}{Dynamic} & \hfil C-band & \hfil 1.5 \\ \cline{3-4}
                     & & \hfil mmWave & \hfil 1.7 \\ \cline{1-4}
                     \hline
                     
                \end{tabular}
            }
                \end{center}
                \label{tab3}
        \end{table}
       
         \begin{figure*}[hbt!]
            \centering
            \subfloat[\centering Static setup in C-band.\label{fig:3a}]{{\includegraphics[trim={0cm 1cm 0cm 4cm}, clip, width=0.34\textwidth]{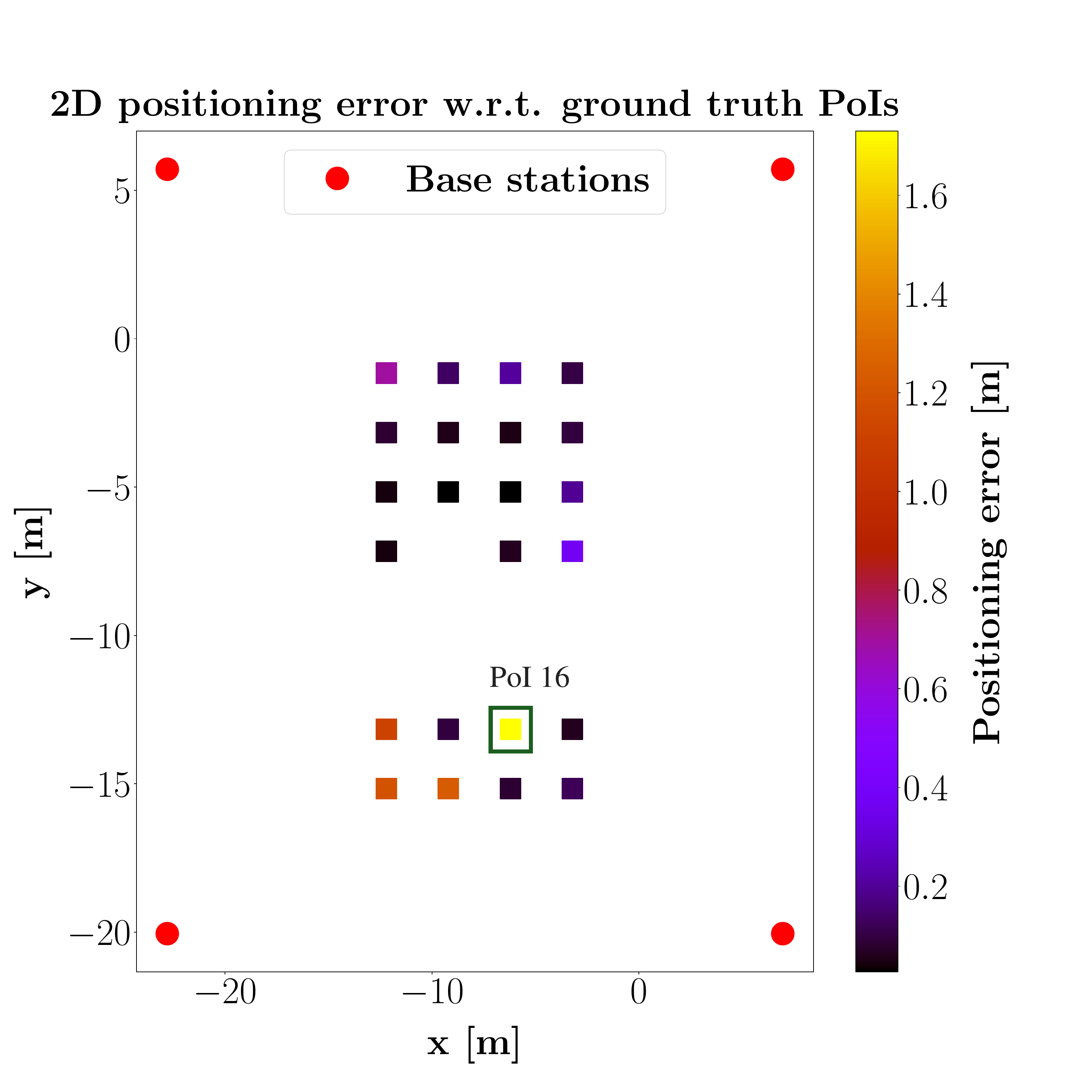} }} 
            \hspace{2mm}
            \subfloat[\centering Static setup in mmWave-band.\label{fig:3b}]{{\includegraphics[trim={0cm 1cm 0cm 4cm}, clip, width=0.34\textwidth]{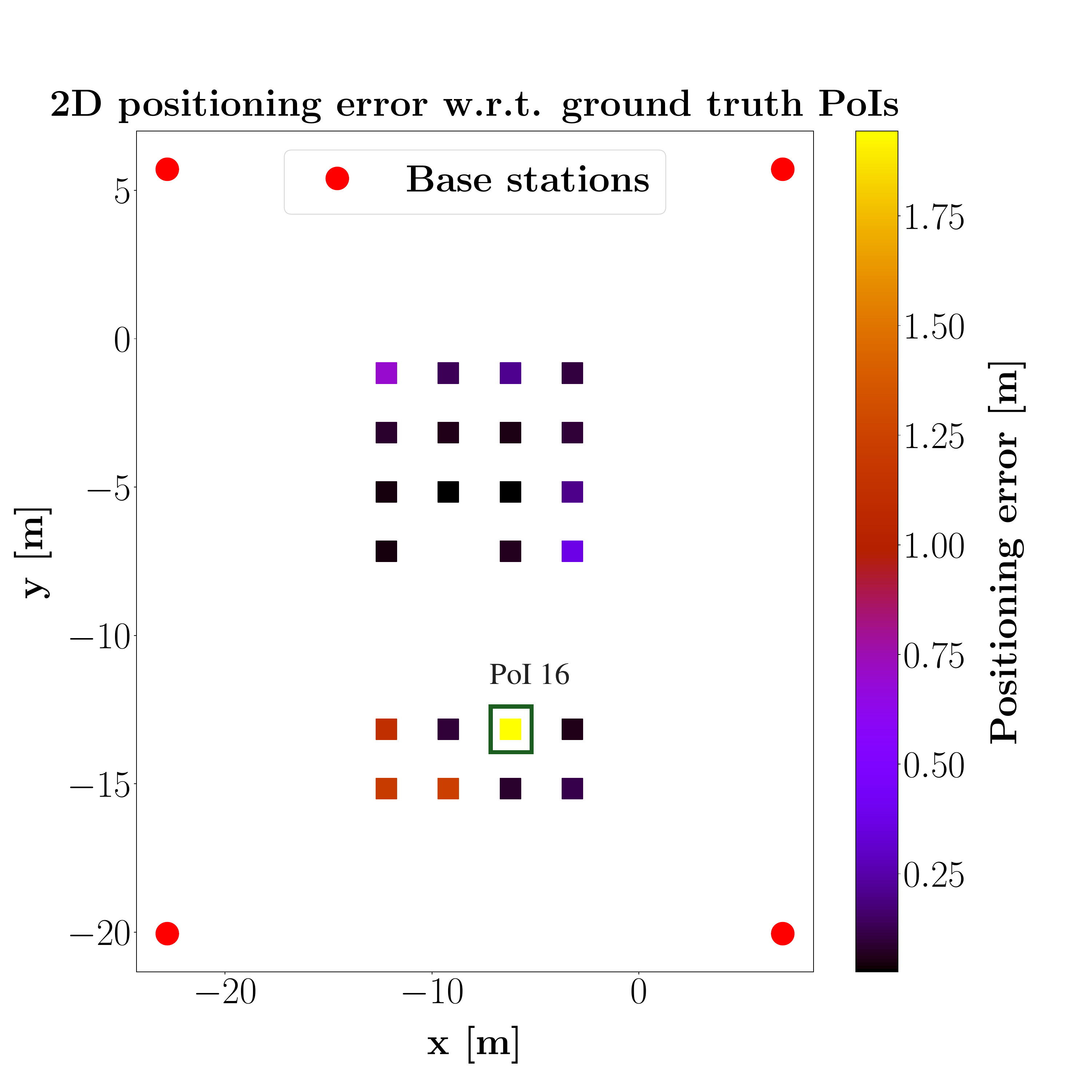} }}%
            \\
             \subfloat[\centering Dynamic setup in C-band.\label{fig:3c}]{{\includegraphics[trim={0cm 1cm 0cm 4cm}, clip, width=0.34\textwidth]{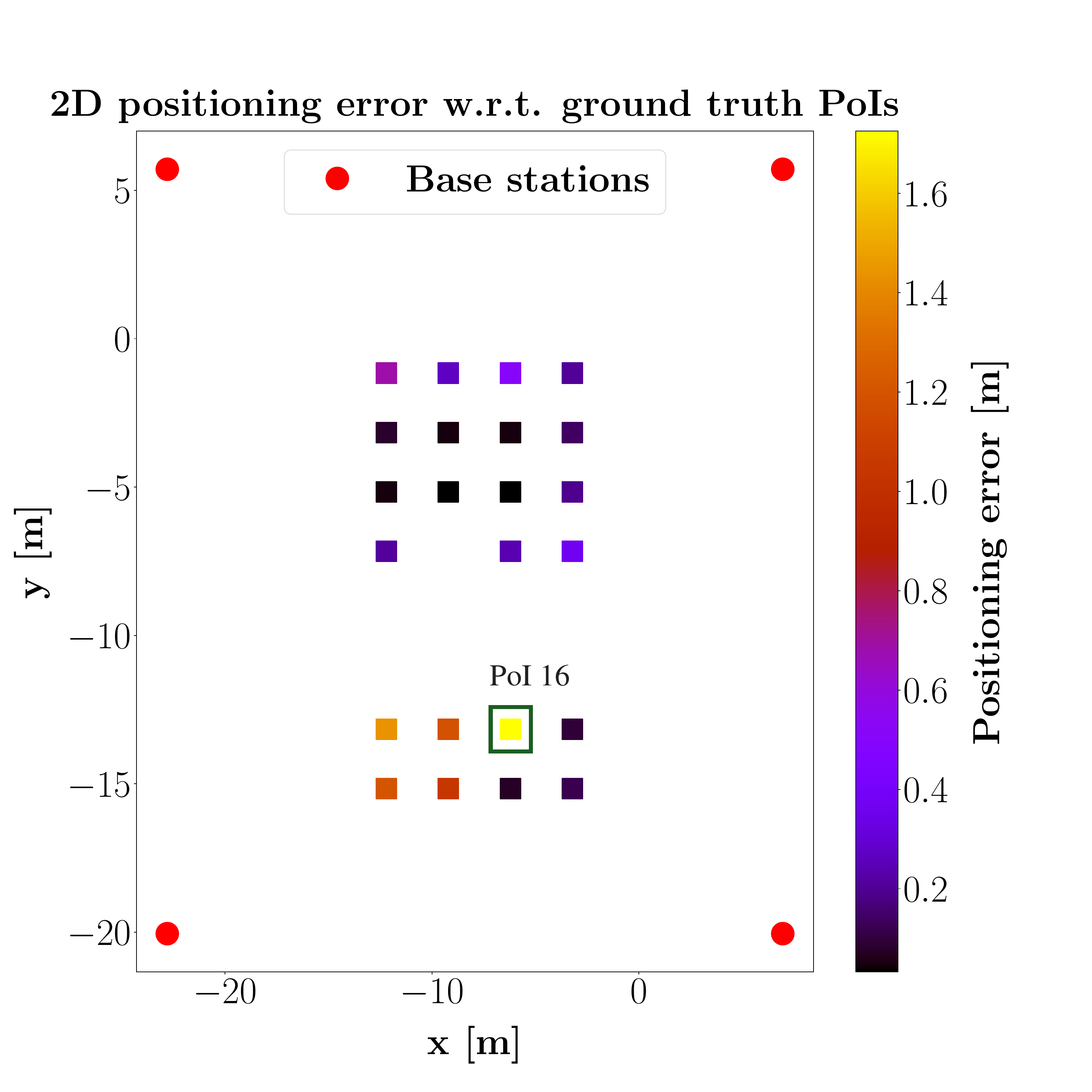} }} 
            \hspace{2mm}
            \subfloat[\centering Dynamic setup in mmWave-band.\label{fig:3d}]{{\includegraphics[trim={0cm 1cm 0cm 4cm}, clip, width=0.34\textwidth]{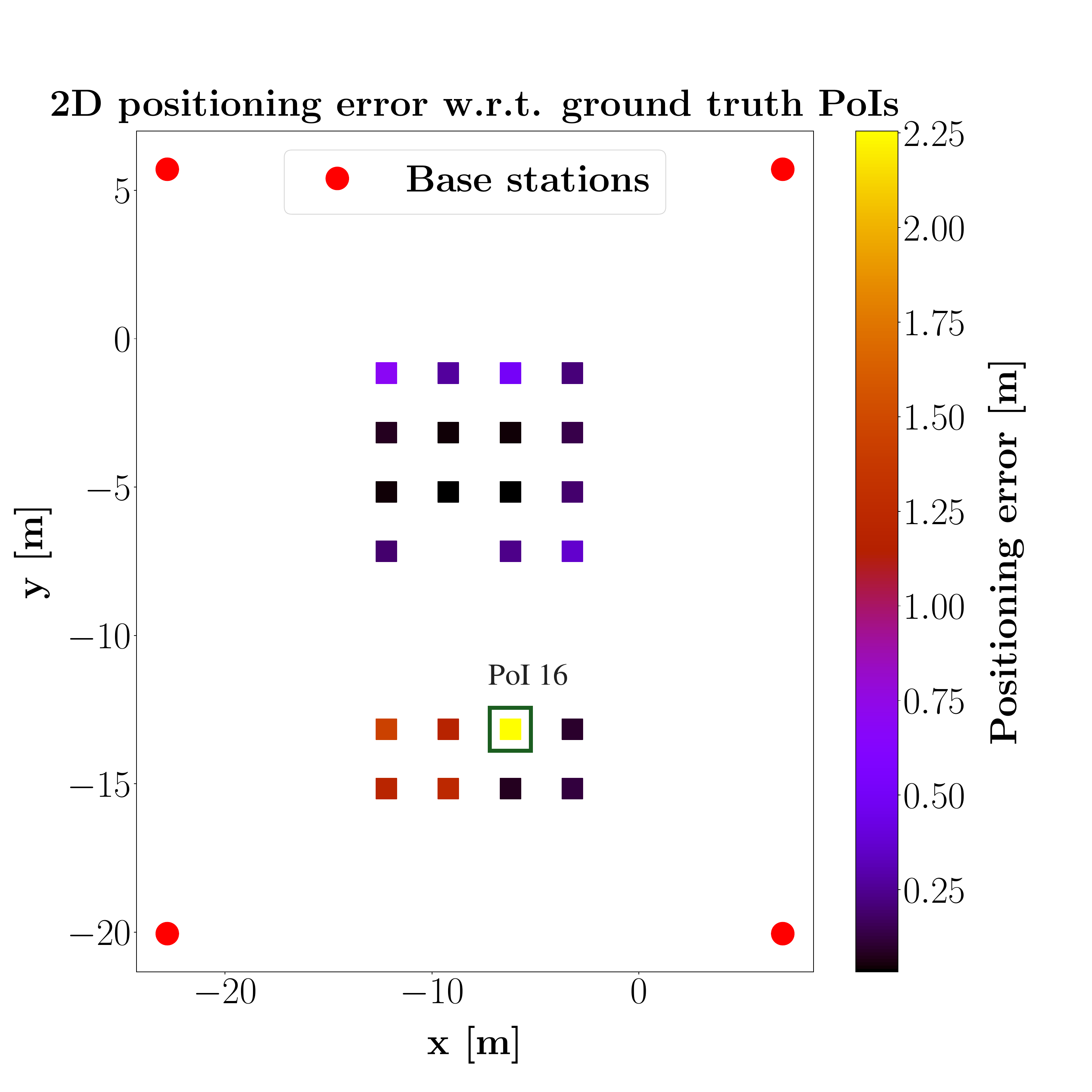} }}%
             \caption{2D positioning error per PoI for the static and dynamic emulation setups.}
            \label{fig:3}
        \end{figure*}
         \begin{figure*}[hbt!]
            \centering
            \subfloat[\centering Static setup in C-band.\label{fig:4a}]{{\includegraphics[height=0.15\textheight,trim={0cm 0cm 0cm 0cm}, clip, width=0.3\textwidth]{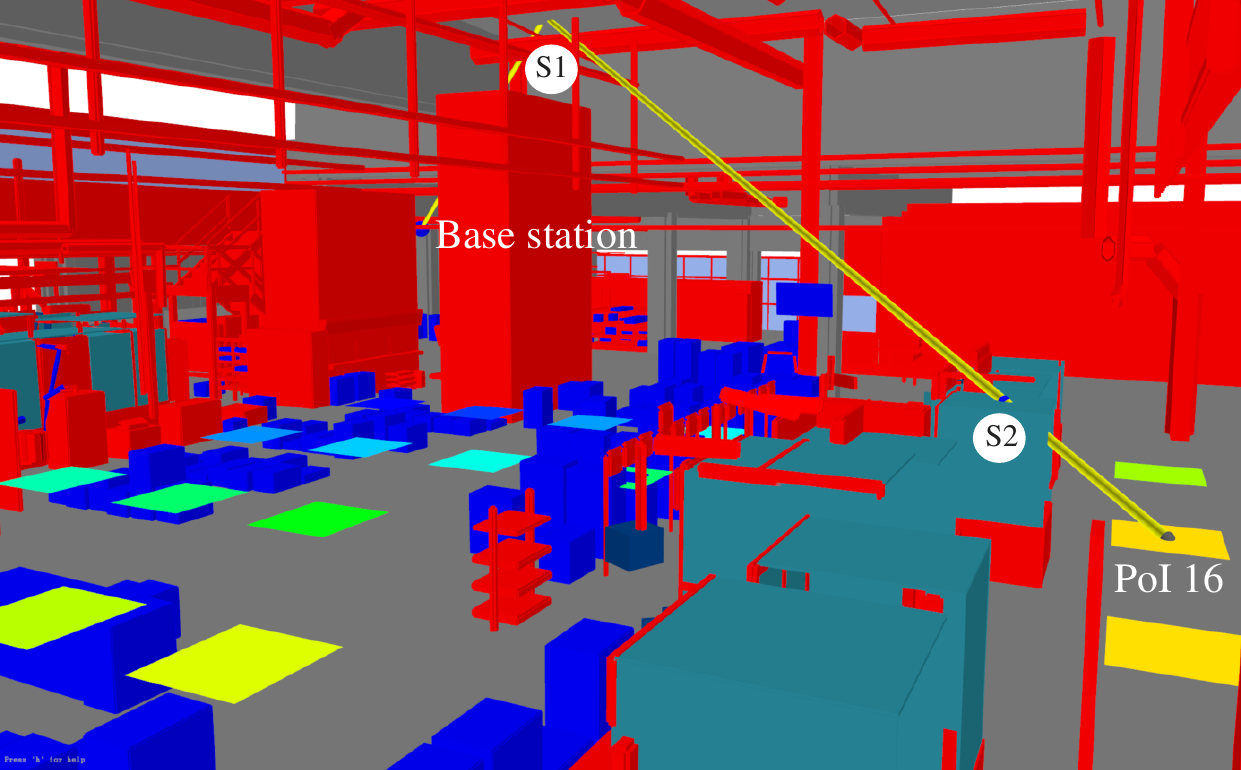} }} 
            \hspace{10mm}
            \subfloat[\centering Static setup in mmWave-band.\label{fig:4b}]{{\includegraphics[height=0.15\textheight, trim={0cm 0cm 0cm 0cm}, clip, width=0.3\textwidth]{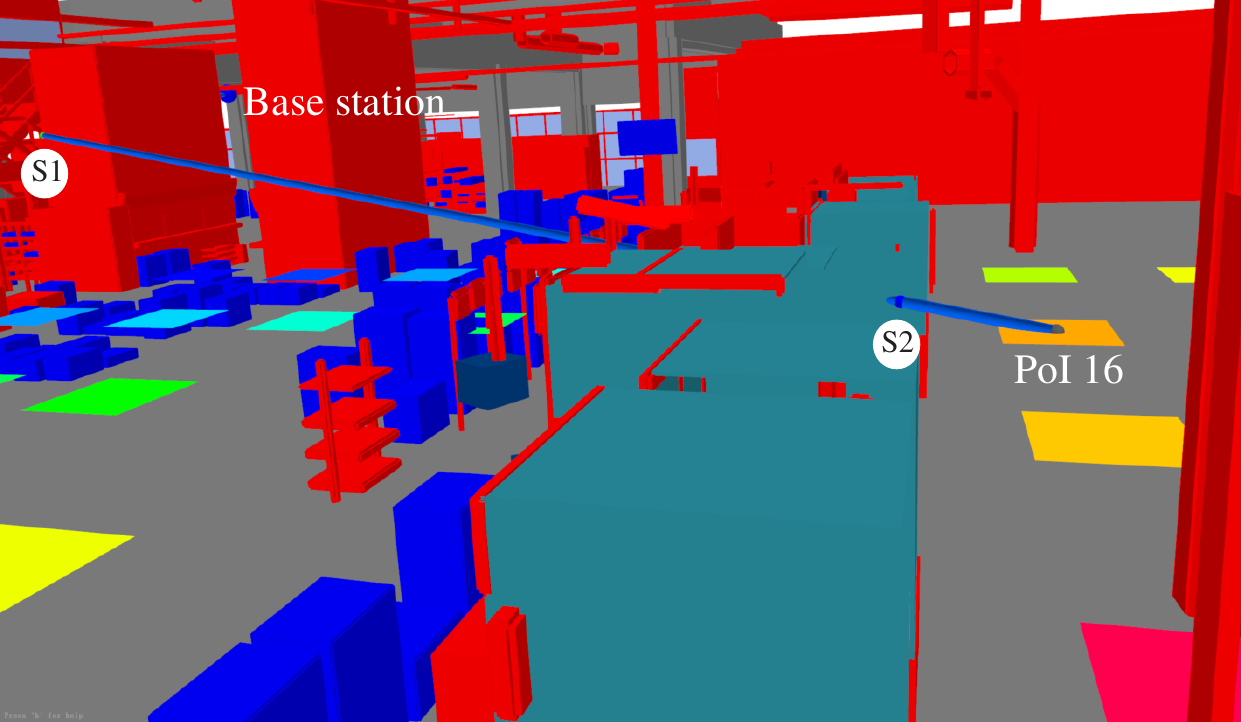} }}
            \\
            \subfloat[\centering Dynamic setup in C-band.\label{fig:4c}]{{\includegraphics[height=0.15\textheight,trim={0cm 0cm 0cm 0cm}, clip, width=0.3\textwidth]{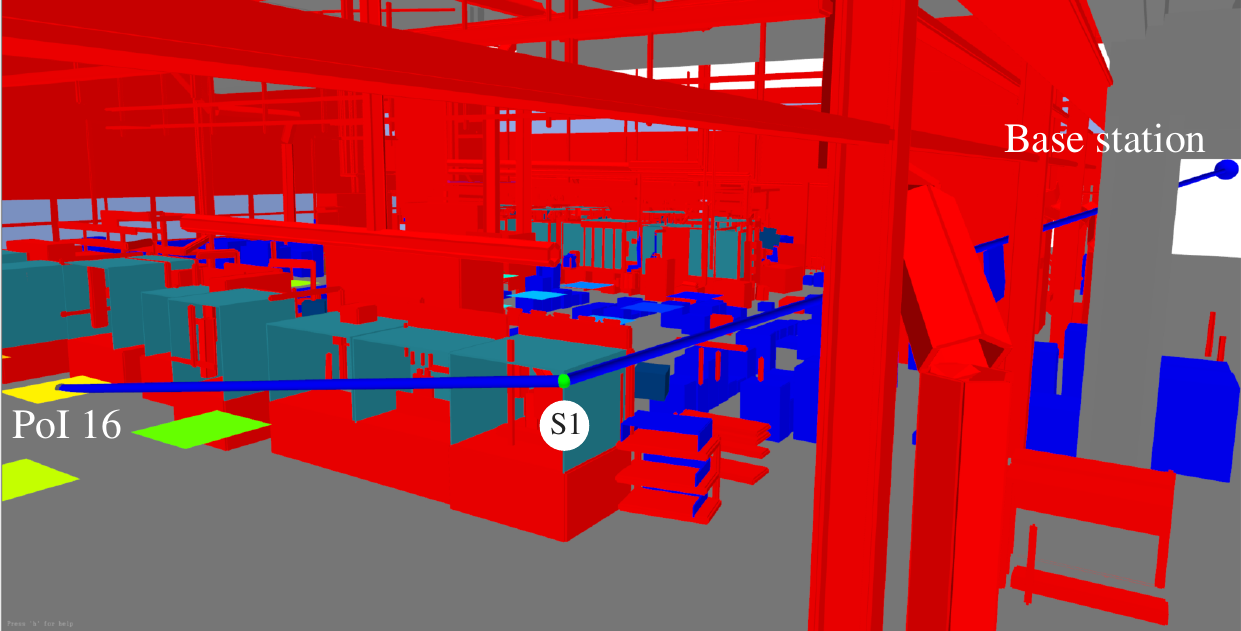} }} 
            \hspace{10mm}
            \subfloat[\centering Dynamic setup in mmWave-band.\label{fig:4d}]{{\includegraphics[height=0.15\textheight,trim={0cm 0cm 0cm 0cm}, clip, width=0.3\textwidth]{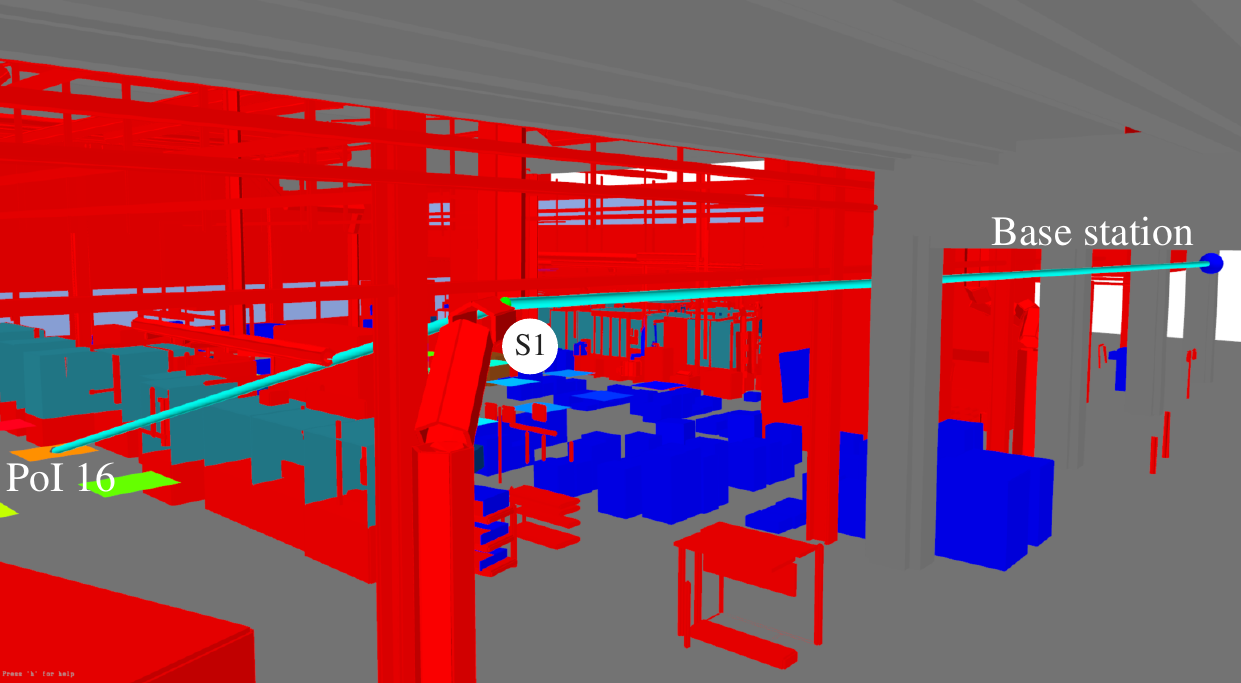} }}%
            \caption{First-arriving MPCs obtained by Ray-Tracing (RT) emulation.}%
            \label{fig:4}
        \end{figure*}

        \subsubsection{Dynamic emulation setup}
        To mimic the reality of the industry, the dynamic behavior in the form of moving forklifts within the production hall is considered in this setup. Analyzing the positioning error per PoI for the current setup, one can note that, compared to the static setup, the positioning error had been increased, even though the BSs and PoIs are not moving. This is because of the fluctuations in the ToA measurements recorded at the UE, due to the NLoS condition caused by the moving forklifts. In addition, looking at the C-band and mmWave-band plots in Fig. \ref{fig:3c} and Fig. \ref{fig:3d}, one can note the difference in positioning error for a few PoIs. For instance, considering the PoI~16, the first-arriving MPC from the BS~3 in the C-band corresponds to the scattered ray with diffraction from the machine indicated by S1 as shown in Fig. \ref{fig:4c}. On the other hand, in mmWave-band for the same BS-PoI pair, the first-arriving MPC corresponds to the scattered ray with diffraction from the column of a building indicated by S1 as shown in Fig. \ref{fig:4d}. The path taken by the diffracted ray in the mmWave-band involving BS-column-PoI covers a larger distance than the path taken by the diffracted ray in C-band involving BS-machine-PoI. As a result, the induced path length error is larger in the mmWave-band, contributing to the increase in positioning error. 
        
        The CDF of 2D positioning error is shown in Fig. \ref{fig:5}. Looking at the curves for the current setup, one can note that similar positioning performance for the PoIs in C-band and mmWave-band can be observed till the 80th percentile of the CDF function, having a positioning error of around 0.9~m. At the 90th percentile, C-band has around 7$\%$ smaller errors than the mmWave-band. Moreover, the C-band continues to outperform the mmWave-band as the positioning error increases. Considering the 90th percentile of the CDF function, we could achieve a positioning error of around 1.5~m and 1.7~m in the C-band and the mmWave-band respectively. Limiting our deployment and analysis to only four BSs, we were able to achieve average positioning errors of 0.43~m and 0.46~m in the C-band and mmWave-band correspondingly.

        \begin{figure}[t]
            \centering
            {{\includegraphics[trim={5cm 0cm 10cm 6.8cm}, clip, width=0.45\textwidth]{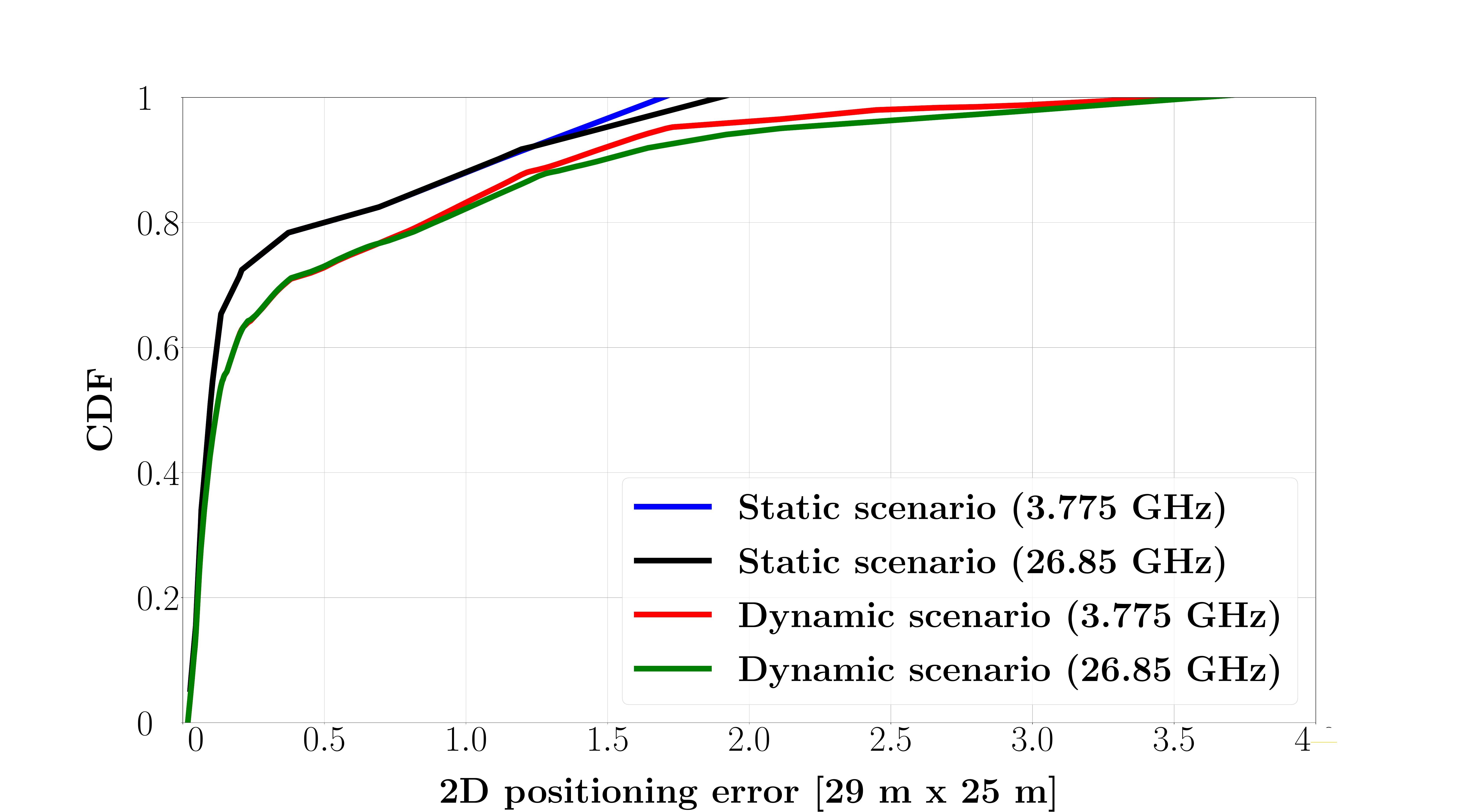} }}
            \caption{Cumulative Distribution Function (CDF) of 2D positioning error for static and dynamic emulation setups.}%
            \label{fig:5}
        \end{figure}


        \section{Conclusion}
        

        In this paper, using an RT engine, we investigated the radio channel in C-band as well as the mmWave-band under static and dynamic emulation setups for a detailed 3D geometric model of an industrial scenario. In addition, we derived the achievable OTDoA-based positioning accuracy in the specific indoor industrial scenario by computing the position of UE using first-arriving MPC. 
        
        The results showed that the diffracted rays are the dominant NLoS MPCs in our emulation setups and have a greater impact on the positioning results. The OTDoA-based positioning meets the 3GPP's release~16 requirements of achieving the positioning error $<$3~m for 80$\%$ of the cases. However, the stringent requirements set by release~17 of achieving positioning error $<$1~m for 90$\%$ of the cases could not be achieved. The outliers in the ToA/distance estimations caused by NLoS condition are the main source of positioning errors. To achieve sub-meter positioning accuracy, new methods for detecting and rejecting outliers beforehand should be implemented. The comparison of indoor positioning performance between the C-band, as well as the mmWave-band, showed that the first-arriving NLoS MPCs used for positioning in mmWave-band are longer in path lengths. A possible explanation for this behavior could be the fact that the higher frequency signals result in higher attenuation when colliding with obstacles. The presence of a higher permittivity obstacle in the path of the first-arriving MPC can prevent it from reaching the PoI. As a result, in mmWave-band, for an MPC to reach the PoI it has to avoid the obstacles with higher permittivity even though the path it takes is longer in distance. Therefore, in our industrial scenario consisting of heavy machinery and concrete walls positioning in the C-band achieves better results compared to positioning in the mmWave-band under both static and dynamic emulation setups. In such a scenario, positioning in mmWave-band requires a high density of BSs to achieve a similar or better accuracy than positioning in C-band. Nevertheless, the extent to which the industries accept the idea of deploying high-density of BSs in indoor scenarios is an open question. 
        



	\section*{Acknowledgment}
	This work has received funding from the European Union's Horizon 2020 research and innovation programme under the Marie Sklodowska-Curie grant agreement ID 956670.

	\bibliographystyle{IEEEtran}
	\bibliography{IEEEabrv,references}
	
\end{document}